\documentclass[aps,prl,twocolumn,superscriptaddress,notitlepage,longbibliography]{revtex4-2}
\usepackage{graphicx}
\usepackage{dcolumn}
\usepackage{bm}
\usepackage{amsfonts}
\usepackage{amssymb}
\usepackage{amsmath}
\usepackage{color}
\usepackage[colorlinks=true,breaklinks=true,linkcolor=blue,
anchorcolor=red,citecolor=blue,urlcolor=blue]{hyperref}
\allowdisplaybreaks[3]

\def\nn{\nonumber}
\def\Eq#1{Eq.~(\ref{#1})}

\def\Fig#1{\text{Fig.}~\ref{#1}}
\def\abs#1{\left|#1\right|}
\def\xk#1{\left(#1\right)}

\newcommand{\De}{\Delta}

\renewcommand{\v}[1]{{\mathbf #1}}

\begin{document}

\title{Nernst Plateau in the Quantum Limit of Low-Carrier-Density Topological Insulators}

\author{Peng-Lu Zhao}
\affiliation{Department of Physics, University of Science and Technology of China, Hefei, Anhui 230026, China}
\affiliation{Quantum Science Center of Guangdong-Hong Kong-Macao Greater Bay Area (Guangdong), Shenzhen 518045, China}
\affiliation{State Key Laboratory of Quantum Functional Materials, Department of Physics, and
Guangdong Basic Research Center of Excellence for Quantum Science, Southern University of Science and Technology (SUSTech), Shenzhen 518055, China}

\author{J. L. Zhang}
\email{zhangjinglei@hmfl.ac.cn}
\affiliation{Anhui Province Key Laboratory of Low-Energy Quantum Materials and Devices, High Magnetic Field Laboratory, HFIPS, Chinese Academy of Sciences, Hefei 230031, China}

\author{Hai-Zhou Lu}
\email{luhz@sustech.edu.cn}
\affiliation{State Key Laboratory of Quantum Functional Materials, Department of Physics, and
Guangdong Basic Research Center of Excellence for Quantum Science, Southern University of Science and Technology (SUSTech), Shenzhen 518055, China}

\affiliation{Quantum Science Center of Guangdong-Hong Kong-Macao Greater Bay Area (Guangdong), Shenzhen 518045, China}


\author{Qian Niu}
\affiliation{Department of Physics, University of Science and Technology of China, Hefei, Anhui 230026, China}
\affiliation{CAS Key Laboratory of Strongly-Coupled Quantum Matter Physics,
University of Science and Technology of China, Hefei, Anhui 230026, China}	

\begin{abstract}
{Nernst effect, a transverse electric current induced by a temperature gradient, is a promising tool for revealing emergent phases of condensed matter. We find a Nernst coefficient plateau in low carrier density topological insulators, as a signature of 1D Weyl points in the quantum limit of the weak topological insulator. The plateau height is inversely proportional to the impurity density, suggesting a way to engineer infinitely large Nernst effects. The Nernst plateau also exists in strong topological insulators, at the bottom of the lowest Landau band. We show that these plateaus have been overlooked in the previous experiments and we highlight the experimental conditions to observe them. Our results may inspire more investigations of employing anomalous Nernst effect to identify emergent phases of condensed matter. 
}
\end{abstract}

\maketitle

\begin{figure}[htbp]
\centering
\hspace{-0.2cm}
\includegraphics[width=0.92\columnwidth]{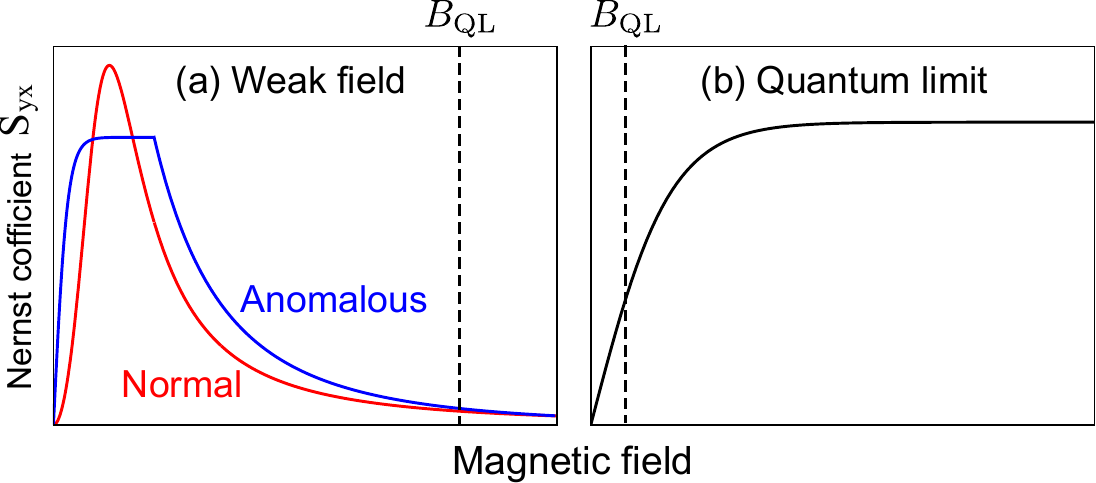}
\caption{(a) Schematic of the previously known two types of the Nernst effect, in terms of the dependence of the Nernst coefficient $S_{yx}$ on the magnetic field at a fixed and low temperature $T$. (b) In comparison, we find a Nernst plateau beyond the expectation of the previous mechanisms,  
at magnetic fields far above the strong-field quantum limit $B_\mathrm{QL}$ (indicated by the black dashed lines), where only the lowest Landau band is occupied by electrons. }\label{FigIntro}
\end{figure}

\emph{Introduction}--Nernst effect is a thermoelectric Hall effect, exhibiting as a transverse electric current generated by a temperature gradient. It is a promising measurement method to reveal outstanding signatures than those in the conductivity.  
Two types of Nernst effects  have been observed in the experiments  \cite{Behnia09JPCM,Behnia16RPP},
the normal one shows a weak-field peak in metals \cite{Mangez76B,Lipc07B,Bel03L,Choi05L,Pourret06L,Kasahara07L,Behnia07L,Spathis08EPL,Pourret07B} and superconductors \cite{Huebener69PR,Pourret06NP,Pourret07B,Spathis08EPL,Tafti14B,Changj12NP, Xuza00N,Wangyy06B} and the anomalous one shows a plateau in topological and magnetic materials in weak fields \cite{Leewl04L,Miyasato07L,Puy08L,Ikhlas17NP,Lixk17L,Lixk18SciP,Sakai18NP,Guin19AM,Xulc20SA,Yanghy20PRM,Wuttke19B} [\Fig{FigIntro}(a)].
A plateau, quantized or not, is always intriguing, thus the Nernst plateau has attracted significant attention in the past ten years. 
It can be explained by either the saturation of magnetization or finite Berry curvature induced by the self-rotation of Bloch wave packets \cite{Changmc95L,Changmc96B,Sundaram99B,Xiaod10RMP}.
Nevertheless, none of these mechanisms applies in strong magnetic fields, where Landau levels are formed and each level produces a quantized Hall conductivity. According to the Mott relation \cite{Smrcka77JPC, Jonson80B, Jonson84B}, both types of the Nernst effect are expected to decay and vanish  in extremely strong magnetic fields. 

In this Letter, however, we theoretically demonstrate that a Nernst plateau can form in the strong-field quantum limit of topological insulators [\Fig{FigIntro}(b)], as
 the ratio of the Nernst coefficient $S_{xy}$ to temperature $T$
 \begin{equation}
	\frac{S_{yx}}{T} =\frac{\pi^2 k_B^2 }{3e} \frac{2}{R \Gamma}, \label{EqSxyI} 
	\end{equation}
where besides the fundamental constants (the Boltzmann constants $k_B$, electron charge $e$, and $\pi$),  
$R=1$ for weak and $R=5$ for strong topological insulators, and 
the Landau-band broadening
$\Gamma$ could remain invariant in the quantum limit, as protected by a detailed charge neutrality resulted from a nontrivial topological phase transition, giving rise to the Nernst plateau. 
Moreover, the height of this Nernst plateau can increase infinitely with decreasing impurity density. Our results not only predict a Nernst plateau in the ultra-quantum limit of low-carrier-density topological insulators, but also suggest a strategy to enhance thermoelectric conversion efficiency.

\emph{Landau bands in topological insulators}--The 3D strong and weak topological insulators can be generically described by the modified Dirac model \cite{Shen17b},
\begin{align}
H_0(\mathbf{k})=&\, \,\hslash v_{x} k_x \tau_z \sigma_x+ \hslash v_{y} k_y \tau_0 \sigma_y+\hslash v_{ z} k_z \tau_x \sigma_x 
 \nn  \\  \label{EqModel} 
 & +\left[\Delta+M_{\perp} \left(k_x^2+ k_y^2\right)+M_z k_z^2\right] \tau_0 \sigma_z,   
\end{align}
where $v_{x,y,z}$ are the Fermi velocities, $\sigma$ and  $\tau$ are Pauli matrices for pseudo and real spins, respectively. $2\abs{\Delta}$ is the bulk gap, $M_{\perp}$ and $M_z$ are two minimal band inversion parameters used to distinguish strong and weak topological insulators \cite{Zhanghj09NP,Chenry15L,Zhangjl19L,Martino19L,Jiangy20L,Wuwb23NM}. Without loss of generality, we assume $\Delta>0$, then \Eq{EqModel} describes a strong topological insulator when $M_z<M_{\perp} <0$ \cite{Ful07PRL,Ful07PRB,Yanbh12L}, a weak topological insulator when $M_{\perp}<0$ and $M_z>0$, or a normal insulator when $M_{\perp}>0$ and $M_z>0$ (see Sec. SI of Supplemental Material \cite{supp}).

By applying a uniform $z$-direction magnetic field magnetic fields $\v B=(0,0,B)$, the energy spectrum of the topological insulator turn to a series of 1D bands of Landau levels. To analyze the Landau bands (details can be found in Sec. SI of Supplemental Material \cite{supp}), the canonical wave vectors are defined by the Peierls substitution $\bm{\Pi}=\bm{k}+e\bm{A}/\hslash$ with the Landau gauge potential $\bm{A}=(-yB,0,0)$. The ladder operators then can be constructed by $\bm{\Pi}_{x}$ and $\bm{\Pi}_{y}$, which further yield the 1D Landau bands as $E_{n \nu}^{\mu}(k_z)=\mu\xk{Z-M_{\perp}/\ell_B^2}+\nu \sqrt{\left(M_zk_z^2+2M_{\perp}n/\ell_B^2+\Delta\right)^2+\left(\sqrt{2n}\hslash v_{\perp}/\ell_B\right)^2}$ for $n\geqslant 1$ and $E_{0 \nu}(k_z)=\nu\abs{M_zk_z^2+M_{\perp}/\ell_B^2+\De-Z}$ for the Lowest Landau bands. Here, $\mu$ and $\nu$ take values of $\pm$, $v_{\perp}=\sqrt{v_xv_y}$, $\ell_{B}=\sqrt{\hslash/eB}$, and $Z=g\mu_B B/2$, where $g$ represents the g-factor and $\mu_B$ stands for the Bohr magneton. $Z$ represents the Zeeman energy, as we have included a Zeeman coupling term $H_{Z}=-g\mu_BB\tau_z\sigma_0/2$ in the free Hamiltonian. The normalized eigenvectors of all Landau bands can be found and are uniformly denoted as $|n\v k,+(\nu),+(\mu)\rangle$.

\begin{figure*}[htb]
\centering
\hspace{-0.2cm}
\includegraphics[width=2.0\columnwidth]{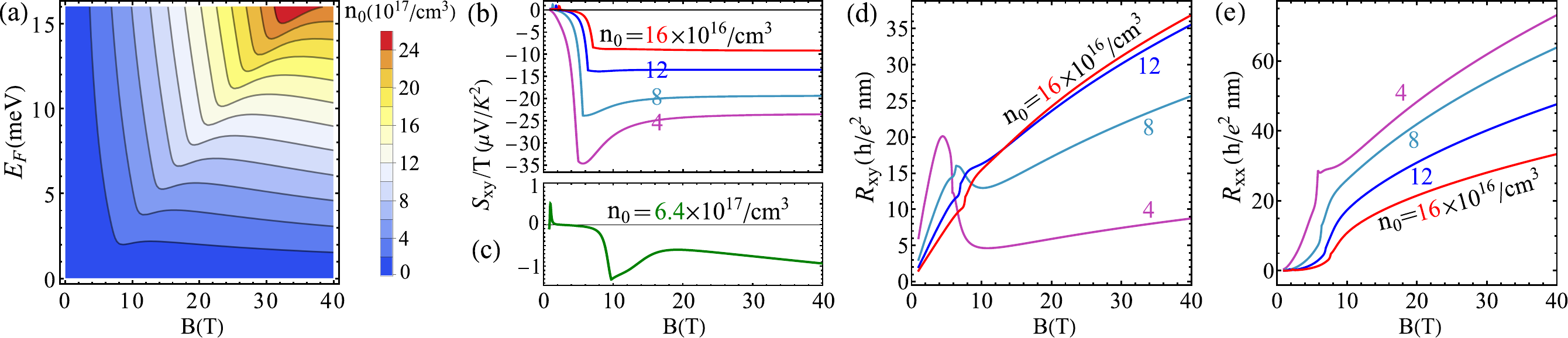}
\caption{Computed functions of $E_F(B)$, $S_{xy}(B)$, $R_{yx}(B)$, and $R_{xx}(B)$ for weak topological insulators at $T=2 \,\mathrm{K}$. (a) Contour plot in the $B-E_F$ plane for a fixed carrier density $n_0$ of weak topological insulators. Each line represents a constant $n_0$, and the different colors correspond to varying values of $n_0$, which are labeled in the right panel. (b), (d), and (e) Calculated $S_{xy}(B)$, $R_{xy}(B)$, and $R_{xx}(B)$, respectively, for different carrier densities labelled by the colored numbers. All plots begin at $B=1\mathrm{T}$--at the quantum limit for $n_0=4\times 10^{16}/\mathrm{cm}^3$, but at the second Landau band for the other three carrier densities. (c) Calculated $S_{xy}(B)$ with an $n_0$ higher than those in (b); in this case, $S_{xy}$ does not exhibit a plateau as in (b). This plot begins in the second Landau band at $6\mathrm{T}$. In all diagrams, we take $M_{\perp}=-12\,\mathrm{eV}\cdot$ \AA$^2$, $M_z= 3\mathrm{eV}\cdot$ \AA$^2$, $\Delta =2.5\,\mathrm{meV}$, $v_{\perp}=5\times 10^5\mathrm{m/s}$, $g=12$, and $\Gamma=2\,\mathrm{meV}$. The quantum limits for $n_0=(4, 8, 12, 16, 64) \times 10^{16}/\mathrm{cm^3}$ are $B_0=0.77, 1.32, 1.82, 2.28, \mathrm{and}\, 6.81 \mathrm{T}$, respectively.  }\label{FigWTI}
\end{figure*}
\begin{figure}[htb]
\centering
\hspace{-0.2cm}
\includegraphics[width=0.98\columnwidth]{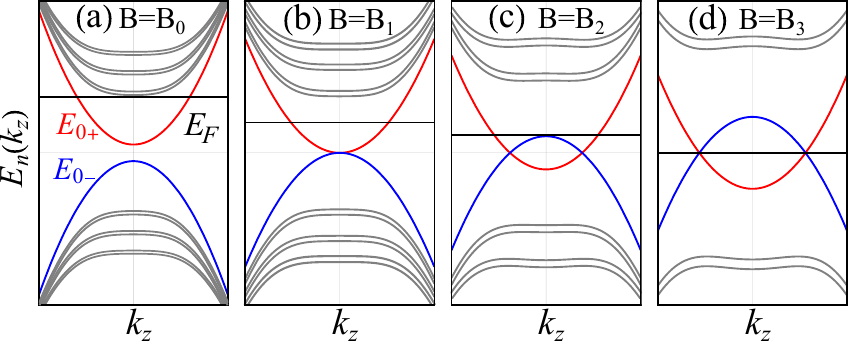}
\caption{Evolution of Landau bands with increasing $B$ at several critical values. The red, blue and gray lines are for the 0th Landau bands $E_{0+}$, $E_{0-}$ and high-index Landau bands, respectively. The black line represents the Fermi energy $E_{F}$. (a) The quantum limit reaches at $B_0$, where the Fermi energy touches the band bottom of $E_{1+}^{-}$. (b) The gap of the 0th Landau band closes at $B_{1}$. (c) The Lifshitz transition happens at $B_{2}$, where $E_{F}$ crosses the bottom of $E_{0-}$. (d) Ideal 1D Weyl states emerge at $B_{3}$, with all Weyl points located at $E_{F}$.}\label{FigWBE}
\end{figure}
\begin{figure*}[htb]
\centering
\hspace{-0.2cm}
\includegraphics[width=2\columnwidth]{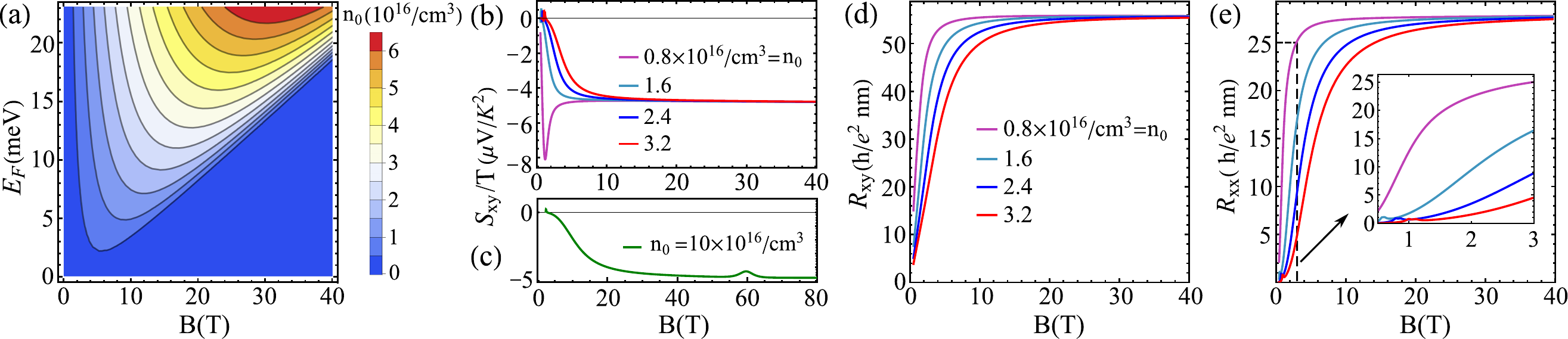}
\caption{Computed functions of $E_F(B)$, $S_{xy}(B)$, $R_{yx}(B)$, and $R_{xx}(B)$ for strong topological insulators at $T=2 \,\mathrm{K}$. (a) Contour plot in the $B-E_F$ plane for a fixed carrier density $n_0$ of strong topological insulators. (b), (d), and (e) Calculated $S_{xy}(B)$, $R_{xy}(B)$, and $R_{xx}(B)$, respectively, for different carrier densities labelled by the colored numbers. All plots begin at $B=0.5\mathrm{T}$--at the quantum limit for $n_0=0.8\times 10^{16}/\mathrm{cm}^3$, but at the second Landau band for the other three carrier densities. The inset in (e) plotted $R_{xx}(B)$ for $B\in \left(0.5, 3\right) \mathrm{T}$, and the labels for the lines in (e) are the same as those in (d). (c) Calculated $S_{xy}(B)$ with an $n_0$ higher than those in (b); in this case, $S_{xy}$ is not flat even for $B \sim 100 \mathrm{T}$. This plot begins in the second Landau band at $3\mathrm{T}$. For all diagrams, we take $M_{\perp}=-120\,\mathrm{eV}\cdot$ \AA$^2$, $M_z= -12\mathrm{eV}\cdot$ \AA$^2$. The values of $\Delta, v_{\perp}, g$, and $\Gamma$ are the same as those used in \Fig{FigWTI}. The quantum limits for $n=(0.8, 1.6, 2.4, 3.2, 10) \times 10^{16}/\mathrm{cm^3}$ are $B_0=0.34, 0.61, 0.86, 1.11, \mathrm{and}\, 3.11 \mathrm{T}$, respectively. }\label{FigSTI}
\end{figure*} 
\begin{figure}[htb]
\centering
\hspace{-0.2cm}
\includegraphics[width=0.95\columnwidth]{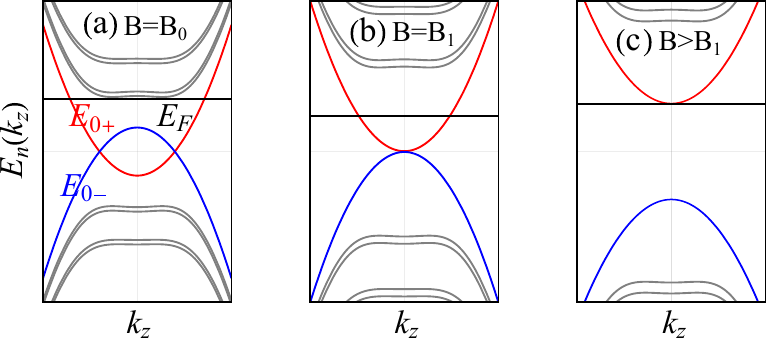}
\caption{Evolution of Landau bands for a strong topological insulator under selected magnetic fields. The labels of the lines are the same as those in \Fig{FigWBE}. (a) The quantum limit reaches at $B_0$. (b) Two Weyl points merge into one at $B_{1}$. (c) The gap between two 0th Landau bands opens after $B_{1}$, and as $B$ increases, $E_F$ gradually pushes to the band bottom.}\label{FigSBE}
\end{figure}

\emph{Conductivities and Nernst coefficient}--In the strong-field quantum limit where only the lowest Landau bands 
cross the Fermi energy, the Nernst coefficient can be found according to the Mott relation, which reads
\begin{align}
S_{x y}=\frac{\pi^2 k_B^2 T}{3 e}\frac{\partial \Theta_{H}}{\partial E_F}, \label{EqSxy}
\end{align}
where the Hall angle $\Theta_{H}\equiv \arctan\left(\sigma_{xy}/\sigma_{xx} \right)$, the longitudinal and Hall conductivities are found by the Kubo-St$\check{\text{r}}$eda formula \cite{Streda82} at low temperature and in the quantum limit (Sec. SIII of Supplemental Material \cite{supp}) as
\begin{align}
\sigma_{x x}= &\,\, \frac{\hbar v_{\perp}^2 e^2}{2 \pi^2\ell_B^2} \int \frac{d k_z}{2 \pi} \sum_{\mu,\nu=\pm} \mathcal{A}_{0\mu}\mathcal{T}_{\mu}^{\nu}  \mathcal{A}_{1\nu}, \label{EqSigmaxx}
\\  
\sigma_{xy}= &\,\, \frac{\hbar v_{\perp}^2 e^2}{2 \pi^2\ell_B^2} \int \frac{d k_z}{2 \pi} \sum_{\mu,\mu'=\pm} \mathcal{T}_{\mu}^{\nu }\left(  \mathcal{A}_{0\mu}\mathcal{G}_{1\nu }-\mathcal{G}_{0\mu}\mathcal{A}_{1\nu } \right), \label{EqSigmaxy}
\end{align}
where $\mathcal{A}_{0\mu}=\Gamma/[\left(E_{0\mu}-E_F\right)^2+\Gamma^2]$ and $\mathcal{A}_{1\nu}=\Gamma/[\left(E_{1\nu}^{\bar{\mu}}-E_F\right)^2+\Gamma^2]$ represent the spectral functions for the $E_{0\mu}$ and $E_{1\nu}^{\bar{\mu}}$ ($\bar{\mu}=-\mu$) bands, respectively, and $\mathcal{T}_{\mu}^{\nu}$ denotes the transition probability between these two bands. Here, $\mathcal{T}_{\mu}^{\nu}(\mu \nu=+)=\left[\sin \left(\theta/2\right) +\gamma \cos \left(\theta/2\right) \right]^2 $ and $\mathcal{T}_{\mu}^{\nu}(\mu \nu=-)=\left[\cos\left(\theta/2\right)-\gamma \sin \left(\theta/2\right)\right]^2 $, where $\gamma=\sqrt{2}M_{\perp}/(\hbar v_{\perp}\ell_B)$ is a dimensionless parameter, and for $\theta \in\left(0,\pi\right)$, $\cos\theta=E_{\mathrm{d}}/\sqrt{E_{\mathrm{d}}^2+\left(\sqrt{2}\hslash v_{\perp}/\ell_B\right)^2} $ with $E_{\mathrm{d} }=M_z k_z^2+2M_{\perp}/\ell_B^2+\Delta$.  $\mathcal{G}_{0\mu}=[\left(E_{0\mu}-E_F\right)^2-\Gamma^2]/[\left(E_{0\mu}-E_F\right)^2+\Gamma^2]^2$ and $\mathcal{G}_{1\nu}=[\left(E_{1\nu}^{\bar{\mu}}-E_F\right)^2-\Gamma^2]/[\left(E_{1\nu}^{\bar{\mu}}-E_F\right)^2+\Gamma^2]^2$. Equations (\ref{EqSigmaxx}) and (\ref{EqSigmaxy}) are  derived under $k_{B}T \ll \Gamma$ to ensure the Mott relation holds. High temperatures and inelastic electron-phonon scattering can break the Mott relation \cite{Smrcka77JPC, Jonson80B, Jonson84B}. For $k_{B}T \ll \Gamma$, phonons are frozen out, inelastic scattering is negligible, and the Mott relation remains valid. Additionally, the quantum limit approximation is well justified when the broadening is much smaller than the Landau band spacing at $k_z=0$, which for typical topological insulators \cite{Zhanghj09NP,Qixl11RMP,Luhz10B} is expressed as $\Gamma\ll \sqrt{2}\hslash v_{\perp}/\ell_B$.

\emph{Nernst plateau in weak topological insulators}--By substituting the calculated conductivities (\Eq{EqSigmaxx} and \Eq{EqSigmaxy}) into the Mott relation (\Eq{EqSxy}), one can obtain the evolution of $S_{xy}$ with respect to $B$. For weak topological insulators, we found that $S_{xy}$ forms a field-independent plateau after a critical value ($\gg B_{\mathrm{QL}}$) (see \Fig{FigWTI}(b)). This behavior manifests under the stringent condition of an extremely low carrier density. 

For a detailed explanation of our findings, we begin by illustrating the evolution of the lowest Landau band with $B$. As shown in \Fig{FigWBE}(a), the two lowest Landau bands have a gap at the quantum limit for a weak topological insulator ($M_{\perp}$, $M_z > 0$, and $\Delta>0$). As $B$ increases, the $E_{0+}$ band shifts downwards while the $E_{0-}$ band shifts upwards. Consequently, the gap between them decreases and closes at $B_1=\De/\left(-M_{\perp}e/\hslash+g\mu_B/2\right)$ (see \Fig{FigWBE}(b)). As $B$ increases further to $B_2$ (see Supplemental Material, Sec. SV \cite{supp} for the calculation of $B_2$),  which is determined by the real root of
\begin{align}
\left(g\mu_B -2M_{\perp}e/\hslash \right) B_2^3 -2\Delta B_2^2-M_z\left( \pi h n_0/e \right)^2= 0, \label{EqCaDeM}
\end{align}
the $E_{0-}$ band reaches the Fermi energy. Here, the carrier density is given by $n_0=\int d k_z  \sum_{n,\mu}\left[f\left(E_{n+}^{\mu}\right)-\left(1-f\left(E_{n-}^{\mu}\right)\right)\right]/4\pi^2\ell_B^2$. Consequently, the Fermi surface in $k_z$-space undergoes a transition from two points to three points, marking a Lifshitz transition \cite{Varlamov85JETP, Varlamov89AP, Blanter94PR}, which is depicted in \Fig{FigWBE}(c). Figure \ref{FigWBE}(d) illustrates that with further increase in $B$, the Fermi energy approaches the Weyl points, forming an ideal 1D Weyl state. Our calculations suggest that such an ideal 1D Weyl state gives rise to the flat $S_{xy}$.

To calculate $S_{xy}(B)$, we need to know how $E_{F}$ changes with $B$. Typically, this can be done by keeping either $E_F$ or the carrier density $n_0$ constant \cite{MahanB,Zhangcl19NC}. When assuming $n_0$ is constant, as shown in \Fig{FigWTI}(a), the calculated $E_{F}$ decreases with increasing $B$ before the Lifshitz transition, after which $E_{F}$ becomes almost field-independent.  Therefore, $S_{xy}$ is calculated by keeping $n_0$ constant before the Lifshitz transition and  $E_{F}$ constant after the transition. The results of $S_{xy}(B)$ at $T=2 \,\mathrm{K}$ are plotted in \Fig{FigWTI}(b), showing that $S_{xy}$ rapidly decays with increasing $B$ before the Lifshitz transition. After the transition, $S_{xy}$ quickly saturates, forming a field-independent plateau. Such a plateau can be analytically obtained as a result of the ideal 1D Weyl state shown in \Fig{FigWBE}(d). Due to charge neutrality at the Weyl point, $\sigma_{xy}=0$, and as a result, 
\begin{align}
S_{x y}=\pi^2 k_B^2 T/3 e\, \sigma_{xx}^{-1} \partial \sigma_{x y}/\partial E_F.\label{EqSxyS}
\end{align}
Moreover, \Eq{EqSigmaxx} and \Eq{EqSigmaxy} in an ideal Weyl state produce (see Supplemental Material, Sec. SVIA\cite{supp})
\begin{align}
	\sigma_{xx}=  \frac{e^2}{\pi h} \frac{\Gamma}{2 M_z k_w},\quad  \frac{\partial \sigma_{xy}}{\partial E_F}=-\frac{e^2}{\pi  h} \frac{1}{M_z k_w},\label{EqRxx} 
\end{align}
where $k_w=\sqrt{\left[\left(-M_{\perp}e/\hslash+g\mu_B/2\right)B-\Delta \right]/M_z}$ representing the position of the Weyl point. Substituting \Eq{EqRxx} into \Eq{EqSxyS} yields $S_{x y}/T =-2\pi^2 k_B^2/(3e\Gamma)$, which corresponds to $R=1$ in \Eq{EqSxyI}. Additionally,  as shown in  Supplemental Material Sec. SVIC \cite{supp}, each subfigure in Fig. \ref{FigWBE} leads to a special point in the Nernst coefficient curve shown in \Fig{FigWTI}(b). Specifically,  Fig. \ref{FigWBE}(a) leads to the peak value of the Nernst coefficient in \Fig{FigWTI}(b); Fig. \ref{FigWBE}(b) leads to the point where the Nernst coefficient crosses zero in \Fig{FigWTI}(b); and Fig. \ref{FigWBE}(c) leads to the valley value in \Fig{FigWTI}(b).

The emergence of this plateau is highly sensitive to the carrier density $n_0$. As shown in \Fig{FigWTI}(c), when $n_0$ is increased to be $6.4\times 10^{17}/\mathrm{cm}^3$, no plateau appears regardless of the magnitude of $B$. The main reason is that for such an $n_0$, the $E_{F}$ after the Lifshitz transition is relatively high (see \Fig{FigWTI}(a)), thus far from the ideal Weyl state. For such case, we have also obtained approximate result of $S_{x y}/T$ under strong magnetic field (see Supplemental Material, Sec. SVI \cite{supp}), which clearly demonstrates the dependence of $S_{xy}$ on $B$ and shows that $S_{xy}$ does not exhibit a plateau in strong magnetic fields. Furthermore, \Eq{EqCaDeM} shows that $B_2 \propto n_0^{2/3}$, and  \Fig{FigWTI}(a) show that as $n_0$ decreases, the Fermi energy at $B_2$ is closer to the 1D Weyl point, meaning $B_3$ decreases with $n_0$.  Therefore, the emergence of a Nernst plateau within experimentally accessible fields requires $n_0$ to be sufficiently low--a condition independent of specific material details--making our prediction highly universal.

The experimental phenomenon of the Nernst plateau bears resemblance to that of the anomalous Nernst effect. To distinguish between them, we also computed the behaviors of the corresponding $R_{xy}(B)$ and $R_{xx}(B)$. Figure \ref{FigWTI}(d) illustrates the variation of $R_{xy}$ with $B$, showing that $R_{xy}$ does not exhibit a plateau. This is a significant distinction between our results and the experimental phenomenon of anomalous Nernst effect, as the plateau in anomalous Nernst effect typically accompanies the anomalous Hall plateau \cite{Xiaod06L,Puy08L,Behnia16RPP}. Furthermore, \Eq{EqRxx}  indicates that when the $S_{xy}$ plateau appears, $R_{xx}$ increases with $B$ according to $\sqrt{\left(-M_{\perp}e/\hslash+g\mu_B/2\right)B-\Delta }$, as shown in \Fig{FigWTI}(e). Observing such behaviors of $R_{xx}$ in experiments provides additional support for validating our plateau theory. 

\emph{Nernst plateau in strong topological insulators}--We will demonstrate that the flat Nernst effect obtained in weak topological insulators also emerges in strong topological insulators (see \Fig{FigSTI}(b)), albeit requiring lower carrier concentrations and involving distinct mechanisms. Firstly, as shown in \Fig{FigSBE}, the evolution of Landau bands in strong topological insulators behaves exactly opposite to that depicted in \Fig{FigWBE} for weak topological insulators. For strong topological insulators ($M_z<M_{\perp} <0$, and $\Delta > 0$), the Weyl points formed by $E_{0+}$ and $E_{0-}$ are present in weak magnetic fields (see \Fig{FigSBE}(a)). As $B$ increases, $E_{0+}$ moves upward, and $E_{0-}$ moves downward. When reaching $B_1$, the two Weyl points merge into one, as shown in \Fig{FigSBE}(b). Figure \Fig{FigSBE}(c) illustrates that with a further increase in $B$, a gap of size $2\left(-M_{\perp}e/\hslash+g\mu_B/2\right)\left(B-B_1\right)$ opens between the two 0th Landau bands.

When calculating $S_{xy}(B)$ for strong topological insulators, we always keep the carrier density $n_0$ constant. If we keep $E_F$ constant, the system will quickly become insulating with increasing $B$ because the $E_{0+}$ band moves upward and the $E_{0-}$ band moves downward. With $n_0$ being constant, as shown in \Fig{FigSTI}(a), the calculated $E_{F}$ first decreases and then increases almost linearly as $B$ increases. Using the calculated $E_{F}(B)$ to compute the $S_{xy}(B)$ yields the results shown in \Fig{FigSTI}(b). Clearly, $S_{xy}(B)$ also flattens in strong magnetic fields. Compared to the case of weak topological insulators shown in \Fig{FigWTI}(b), the carrier density here is nearly an order of magnitude lower, and while the Nernst plateau values vary with carrier density in weak topological insulators, they remain constant in this case. The flat $S_{xy}$ appeared here can be understood as the effect of the band bottom shown in \Fig{FigSBE}(c), which is completely different from the ideal Weyl state in the case of weak topological insulators. Near the band bottom, $\sigma_{xy}$ is comparable to $\sigma_{xx}$. To obtain the approximate results of the Nernst coefficient, we need to calculate the asymptotic results of $\sigma_{xy}$, $\sigma_{xx}$, $\partial \sigma_{xy}/\partial E_F$ and $\partial \sigma_{xx}/\partial E_F$ separately. The detailed calculations are presented in Supplemental Material, Sec. SVIB \cite{supp}, and the approximate results are as follows:
\begin{align}
\sigma_{xx}=  \frac{e^2}{\pi h} \frac{ \sqrt{\Gamma  }}{ 4\sqrt{2 \abs{M_z}} },	\,\,	\frac{\partial \sigma_{xx}}{\partial E_F}=\frac{e^2}{\pi  h}  \frac{1}{ 8\sqrt{2 \abs{M_z} \Gamma } }, \label{EqSigxxS}
\\ 
\sigma_{xy}=  -\frac{e^2}{\pi h} \frac{ \sqrt{\Gamma  }}{ 2\sqrt{2 \abs{M_z} } }, \,\,	\frac{\partial \sigma_{xy}}{\partial E_F}=-\frac{e^2}{\pi  h}  \frac{3}{ 4\sqrt{2 \abs{M_z} \Gamma } }.  \label{EqSigyx}
\end{align}
The asymptotic value of the Nernst coefficient is
$S_{x y}/T =-2\pi^2 k_B^2/(15e\Gamma)$, i.e., the result for $R=5$ in \Eq{EqSxyI}, which exactly gives the value of the plateau in \Fig{FigSTI}(b) when $\Gamma= 2\,\mathrm{meV}$.
Experimentally observing the plateau requires a low density. As shown in \Fig{FigSTI}(c), when $n=6.4\times 10^{16}/\mathrm{cm^3}$, no plateau appears even at magnetic fields as high as $80\mathrm{T}$, a value exceeds the limit of steady magnetic fields achievable in current experiments.

To distinguish the flat Nernst effect in strong topological insulators from the case in weak topological insulators and the anomalous Nernst behavior, we calculated $R_{xy}(B)$ and $R_{xx}(B)$, which are shown in \Fig{FigSTI}(d) and \Fig{FigSTI}(e), respectively. Clearly, $R_{xy}(B)$ and $R_{xx}(B)$ also become flat under strong magnetic fields, and the values of the plateaus can be obtained through $\sigma_{xx}$ and $\sigma_{xy}$ in \Eq{EqSigxxS} and \Eq{EqSigyx}, respectively. Simultaneous saturation of $R_{xy}(B)$ and $R_{xx}(B)$ is rare in topological insulators but was recently observed in $\mathrm{ZrTe}_5$ \cite{Gourgout22QM, Shih25NT}.

\emph{Experimental Observations}--So far, only a few experiments have observed the Nernst plateau in the quantum limit. Recent experiments in $\mathrm{HfTe}_5$ \cite{Zhangy24arXiv} have observed Nernst plateaus in magnetic fields ranging from 15T to 32T ($B_{\mathrm{QL}}\sim 1.5\mathrm{T}$), and the measured $R_{xx}$ fits well with $\sqrt{B}$ when the Nernst plateau appears. Given that recent infrared magneto-optical experiments have identified $\mathrm{HfTe}_5$ as a weak topological insulator \cite{Wuwb23NM}, and that our model \Eq{EqModel} and the parameters in \Fig{FigWTI} also apply to  $\mathrm{HfTe}_5$, we believe that the experimental results in \cite{Zhangy24arXiv} support our Nernst plateau theory for weak topological insulators. For the case of strong topological insulators, we think that the Nernst plateau observed earlier in $\mathrm{ZrTe}_5$ \cite{Liang18NP} can be explained by our theory. First, our model \Eq{EqModel} and the parameters in \Fig{FigSTI} apply to $\mathrm{ZrTe}_5$ \cite{Chenry15L,Tangfd19N}. Second, the carrier density (hole) is very low, leading to $B_{\mathrm{QL}}\sim 0.3\mathrm{T}$ \cite{Liang18NP,Wanghc18SA, Shahi18PRX}, while the observed Nernst plateau appears in the 3–6T range, which is well beyond the quantum limit.

We anticipate that more experiments reporting this phenomenon in other typical topological insulators such as $\mathrm{Bi_{2}Se_{3}}$, $\mathrm{Bi_{2}Te_{3}}$, and $\mathrm{Sb_{2}Te_{3}}$\cite{Zhanghj09NP,Qixl11RMP,Luhz10B}. For these materials, the Hamiltonian includes an additional term, $ (C+D_zk_z^2+D_{\perp}k_{\perp}^2 ) \mathrm{I}_{4\times 4}$. This term breaks the electron-hole symmetry, resulting in different masses for the lowest electron and hole Landau bands, but does not affect the existence of the Nernst plateau given by \Eq{EqSxyI} (see Supplemental Material, Sec. SVIII \cite{supp}). However, the challenge in observing the Nernst plateau in these materials is engineering a sufficiently low $n_0$.

\emph{Discussion}--In our calculations, we assumed $\Gamma$ to be field-independent, as supported by previous studies \cite{Gornik85L, Smith85B, Ashoori92SSC, Wangcj21L}. Since the Nernst plateau value is proportional to $1/\Gamma$, we confirmed $\Gamma$ remains field-independent by calculating it using the Born approximation with Gaussian disorder \cite{Abrikosov98B,Luhz15B,Luhz15B2,Zhangsb16njp}. Details are in the Supplemental Material, Sec. SVII \cite{supp}. The results show that both strong and weak topological insulators can have a field-independent $\Gamma$ when the Nernst plateau appears. The Nernst plateau value is approximately inversely proportional to impurity density. Reducing the impurity density allows the Nernst plateau value predicted by our theory to surpass the current record, albeit constrained by an upper limit (see Supplemental Material, Sec. SVIIC \cite{supp}), thereby significantly improving thermoelectric conversion efficiency.

\emph{Acknowledgments}--We thank Yang Gao, Jian-hui Zhou, Zhi Wang and Song-bo Zhang for fruitful discussions. The work is supported by the National Key R\&D Program of China (Grants No. 2023YFA1406300, No. 2022YFA1403700 and No. 2022YFA1602602), the National Natural Science Foundation of China (Grants No. 12304074, No. 12234017, No. 12374041, No.12525401, No.12474053, and No. 12350402), Guangdong Basic and Applied Basic Research Foundation (Grant No. 2023B0303000011), Guangdong Provincial Quantum Science Strategic Initiative (Grants No. GDZX2201001 and No. GDZX2401001), the Science, Technology and Innovation Commission of Shenzhen Municipality (Grant No.ZDSYS20190902092905285), High-level Special Funds (Grant No. G03050K004), the New Cornerstone Science Foundation through the XPLORER PRIZE, and the Basic Research Program of the Chinese Academy of Sciences Based on Major Scientific Infrastructures (Grant No. JZHKYPT-2021-08). The numerical calculations were supported by the Center for Computational Science and Engineering of SUSTech.

\emph{Data availability}--No data were created or analyzed in this article.

\bibliographystyle{apsrev4-2allauthors}
\bibliography{NernstPlateau.bib}

\end{document}